\documentclass[a4paper,fleqn,usenatbib]{mnras}

\usepackage{newtxtext,newtxmath}
\usepackage[T1]{fontenc}
\usepackage{ae,aecompl}

\usepackage{graphicx}	% Including figure files
\usepackage{amsmath}	% Advanced maths commands
\usepackage{amssymb}	% Extra maths symbols

\title[Probing the ionosphere with RadioAstron] 
{Probing the ionosphere by the pulsar B0950+08 with help of RadioAstron ground-space baselines}

\author[V. I. Zhuravlev et al.]{
Vladimir I. Zhuravlev,$^{1}$\thanks{E-mail: zhur@asc.rssi.ru (VIZ)}
Yu. I. Yermolaev,$^{2}$
A. S. Andrianov$^{1}$
\\
$^{1}$Lebedev Physical Institute, Russian Academy of Sciences, Profsoyuznaya str. 84/32, Moscow 117997, Russia\\
$^{2}$Space Research Institute, Russian Academy of Sciences, Profsoyuznaya str. 84/32, Moscow 117997, Russia
}

\date{Accepted XXX. Received YYY; in original form ZZZ}

\pubyear{2015}

\begin{document}
\label{firstpage}
\pagerange{\pageref{firstpage}--\pageref{lastpage}}
\maketitle

% Abstract of the paper
\begin{abstract}
The ionospheric scattering of pulses emitted by PSR B0950+08 is  measured using the 10-m {\it RadioAstron Space Radio Telescope}, the 300-m Arecibo Radio Telescope  and  the 14x25-m Westerbork Synthesis Radio Telescope (WSRT)  at  a frequency band  between 316  and 332 MHz. We analyse this phenomenon based on a simulated model of the phase difference obtained between antennas that are widely separated  by nearly 25 Earth diameters. We present a technique for processing and analysing  the ionospheric total electron content (TEC) at the ground stations of the ground-space interferometer. This technique allows us to derive almost synchronous half-hour structures of  the TEC  in the ionosphere at an intercontinental distance between the Arecibo and  WSRT stations. We find that  the amplitude values of the detected structures are approximately  twice as large  as the values for the TEC  derived in the International Reference Ionosphere (IRI) project. Furthermore,  the values of the TEC outside these structures are almost the same as the corresponding values found by the IRI. According to a preliminary analysis, the detected structures  were observed during a geomagnetic storm with a minimum Dst index of $\sim 75$\,nT generated by interplanetary disturbances, and  may be due to the influence of interplanetary and magnetospheric phenomena on ionospheric disturbances. We show that the Space Very Long Baseline Interferometry  provides us  with new opportunities to study the TEC, and we demonstrate the capabilities of this instrument to research the ionosphere.
\end{abstract}


%%%%%%%%%%%%%%%%%%%%%%%%%%%%%%%%%%%%%%%%%%%%%%%%%%

%%%%%%%%%%%%%%%%% BODY OF PAPER %%%%%%%%%%%%%%%%%%

\section{Introduction}
\label{sec:introduction}
The scattering of signals from compact radio sources such as pulsars can be used to study small-scale fluctuations  in the electron density of intervening turbulent plasmas. Propagation effects are most clearly accompanied by intensity modulation in the frequency and time domains, pulse and angular broadening with decreasing frequency and signal dispersion. We researched plasma inhomogeneities in the directions of several pulsars using ground-space Very Long Baseline Interferometry (VLBI) observations. These  observations were conducted using projected baselines of up to 220,000\,km, thus providing the highest angular resolution ever achieved at metre wavelengths for the detailed study of distant objects in the Galaxy. Starting with the first observations of pulsars B0950+08 \citep{2014ApJ...786..115S} and B0329+54 \citep{2016ApJ...822...96G}, which were carried out within the framework of the {\it RadioAstron  Early Science Program} \citep{2012evn..confE.109S}, it became clear that most of the measured parameters cannot be obtained solely at ground level. Our analysis was based on the fundamental behaviour of structure and coherence functions. It is shown that the scintillation of the signal has a modulation of less than $100\%$ at frequencies that are much less than the carrier frequency. This  modulation could be caused by scattering layers and a "cosmic prism" , i.e. a sufficiently strong gradient in the column density of refracting material that causes changes in phase located between a  pulsar and Earth. According to our estimation, the Fresnel zone of refractive radiation of both pulsars exceeds one Earth diameter. Ground-space interferometry enables us to measure this scale directly, rather than inferring it via the movement of the pulsar, scattering material and single antenna. Technical information on the {\it RadioAstron mission} can be found in \citet{2013ARep...57..153K,2014CosRe..52..319A}.

Radio waves of cosmic origin received at ground stations always pass through the Earth's ionosphere, and are recorded at ground VLBI stations with an additional delay. The total electron content (TEC) is an important ionospheric parameter in the propagation of radio waves  and is used in other fields such as space geodesy, spacecraft navigation, and communication. The state of the ionosphere is mainly affected by the Sun via photo-ionisation under UV radiation and soft X-rays, injection of particles from the solar wind and ionisation of the atoms and molecules of the ionosphere by cosmic rays. Unusual phenomena in the ionosphere are associated with the action of particles and radiation during solar or magnetospheric flare events. 

Ionospheric effects are a source of phase errors in the visibilities due to destruction of the phase coherence of the signal between two elements of an interferometer. The cross-power spectrum or cross-spectrum $\tilde{S}(\vec{\boldsymbol r}_1,\vec{\boldsymbol r}_2;\nu,t)$ at a frequency $\nu$  is the conjugate multiplication of the complex electric field at two antennas:
\begin{eqnarray}
\tilde{S}(\vec{\boldsymbol r}_1,\vec{\boldsymbol r}_2;\nu,t)=\tilde{S}(\vec{\boldsymbol r}_1;\nu,t)\tilde{S}^*(\vec{\boldsymbol r}_2;\nu,t)
\label{eq:0}
\end{eqnarray}
where $\vec{\boldsymbol r}_1$ and $\vec{\boldsymbol r}_2$ are the spatial coordinates in the plane of the observer perpendicular to the line of sight. $\tilde{S}(\vec{\boldsymbol r}_1,\vec{\boldsymbol r}_2;\nu,t)$ is known as the visibility in the frequency domain. The possibility of the change in the cross-spectrum over time, in increments longer than the accumulation time of the single spectrum is determined by the argument $t$. This is known as a dynamic cross-spectrum. After propagation through a turbulent medium, the spectrum of the electric field for one antenna is represented as:
\begin{eqnarray}
\tilde{S}(\vec{\boldsymbol r}_1;\nu,t)=\tilde{s}(\vec{\boldsymbol r}_1;\nu,t)\exp [-i\Psi (\vec{\boldsymbol r}_1;\nu,t)]
\label{eq:0_2}
\end{eqnarray}
where $\tilde{s}(\vec{\boldsymbol r}_1;\nu,t)$ is the modulation coefficient, and $\Psi (\vec{\boldsymbol r}_1;\nu,t)$ is the phase delay determined by the ionosphere and the "cosmic prism" .  An inverse Fourier transform  $\mathfrak{F}^{-1}$ of $\nu$ to the delay $\tau$ leads to the cross-correlation function  of electric fields:
\begin{eqnarray}
S(\vec{\boldsymbol r}_1,\vec{\boldsymbol r}_2;\tau,t)=\mathfrak{F}^{-1}\bigl[\tilde{S}(\vec{\boldsymbol r}_1,\vec{\boldsymbol r}_2;\nu\to\tau,t)\bigr]
\label{eq:0_1}
\end{eqnarray}
where $S(\vec{\boldsymbol r}_1,\vec{\boldsymbol r}_2;\tau,t)$ is known as the visibility  in the delay domain.

When radio waves pass though the Earth's ionosphere, the majority of the phase delay in the  signals is caused by the F2 layer, which has the highest concentration of charged particles. A radio sounding of the ionosphere enables us to derive information about the plasma structures of the TEC along the path of the radio signal. The results of such observations are discussed in \citet{2006JGRA..111.7S11H}, which identifies medium-scale travelling ionospheric disturbances (MSTIDs) with periods lower than 20\,min, and large-scale travelling ionospheric disturbances (LSTIDs) with a period greater than 1\,h. These are also described in \citet{2012RaSc...47.0L02H}, where ionospheric fluctuations are interpreted as the sum of two turbulent components with maximum scales of about 10\,km and 300\,km,  and in \citet{1992A&A...257..401J}, where Very Large Array (VLA) phase-baseline data are used to evaluate ionospheric plasma irregularities. A more sophisticated method for analysing simultaneous VLA and optical observations is proposed in \citet{2009RaSc...44.0A11C}, which gives an improved understanding of dominant, night-time summer ionospheric phenomena contributing to VLA signal distortion in great detail. The use of polarised signals travelling through the ionosphere enables us to derive information about the distribution of the magnetic field via an analysis of the Faraday rotation of the plane polarisation of the signal \citep[e.g.,][]{1970pewp.book.....G}. The first attempt to evaluate the influence of the ionosphere on ground-space observations was made in \citet{2017MNRAS.468.3709S}. The authors found periodic fluctuations in a complex cross-spectrum with period 70\,s, phase-shifted by 90\degr, for the Green Bank-Westerbork baseline. These fluctuations were due to changes in the ionosphere over time. It was shown that the response of the ground interferometer was more strongly distorted by the ionospheric phase than the corresponding response of the ground-space interferometer. These results were taken into account when the interstellar refractive component was analysed. 

There has been increasing interest of late in the ionosphere, which has been stimulated by a new generation of wide-field low-frequency instruments such as the Low-Frequency Array (LOFAR, 10--250\,MHz, \citealt{2013A&A...556A...2V}), the Long Wavelength Array (LWA, 10--80\,MHz, \citealt{2009IEEEP..97.1421E}), and the Murchison Widefield Array (MWA, 80--300\,MHz, \citealt{2013PASA...30....7T}). Experimental data related to the inhomogeneities in the ionosphere have been obtained mainly using radiophysical methods such as ground-based soundings of the ionosphere and measurements of radiation from cosmic sources and signals from artificial Earth satellites to ground tracking stations. The Global Positioning System (GPS) and Russian global navigation system (GLONASS) at two frequencies have been used  to simulate the plasma effects of the ionosphere in radio astronomical observations \citep{2003AnGeo..21.2083M,2006JGRA..111.7S11H,2016PASA...33...31A}. Dual-frequency dispersive phase measurements  such as these allow us to estimate the TEC between a GPS/GLONASS receiver and a broadcasting satellite. The GPS system transmits dual-frequency data at central frequencies L1=1575.42 MHz and L2=1227.6 MHz. To enable clear differentiation from GPS, the GLONASS carrier frequencies are denoted using G instead of L. The central frequency G1 is 1602.0 MHz and G2 is 1246.0 MHz \citep{Tsui2000,Hofmann2008}.

In geodetic VLBI, observations are performed at two frequencies 2.3\,GHz (S band) and 8.4\,GHz (X band) to correct for the ionospheric delay \citep{2003RaSc...38.1069S,2006RaSc...41.1006H}. The dispersive properties of the ionosphere could be used in combination with independent GPS-based TEC. These measurements are made at stations near to VLBI sites. In \citet{2000A&A...356..357R}, accuracies of up to sub-milli-arc-second levels were achieved  at 8.4\,GHz using GPS TEC. It was shown that the GPS and VLBI values for  ionospheric delay agreed with root-mean-square differences of below 0.15\,ns for intercontinental baselines, and 0.10\,ns for continental ones. This 0.1--0.15\,ns delay corresponds to 5.3--7.9\,TECU at 8.4\,GHz, where a TECU is the typical vertical value at night (1 TECU = $10^{16}$\,electrons/m$^2$). For our ground-space experiment at 324\,MHz, an uncertainty of 5.3--7.9\,TECU corresponds to a delay error of $\tau_{err}$=70--100\,ns (see Equation~\ref{eq:12} below and $\tau_{err}=\varphi/2\pi\nu$). Although a temporal correction of this size is acceptable in VLBI observations  at 324\,MHz, an ionospheric analysis requires more accurate TEC data.

The {\it RadioAstron} project provides users with another opportunity to study the ionosphere at the ranges of large baselines. If the orbital telescope is located at a sufficient distance from the Earth's surface, then only the spectra detected at the ground VLBI telescopes will include modulations caused by the ionosphere, since emissions from a cosmic source to the orbital antenna do not pass through the Earth's ionosphere. It then becomes possible to determine the absolute value of the TEC along the line of sight from the source to the ground station. 

Pulsars are a good tool for investigating integrated ionospheric parameters, due to their very small angular sizes and high levels of coherence. We chose one of the closest  pulsars, PSR B0950+08  for this study from the pulsar catalogue ATNF\footnote{http://www.atnf.csiro.au/people/pulsar/psrcat/} \citep{2005AJ....129.1993M}, which has a relatively high flux density at metre wavelengths. The annual parallax measurement using VLBI gives a distance of 0.26\,kpc \citep{2017ApJ...835...29Y}. Its barycentric period is $P_{\circ}=0.253065$\,s \citep{2004MNRAS.353.1311H}, and its dispersion measure is DM=2.96927\,pc/cm$^3$ \citep{2015ApJ...808..156S}. This pulsar lies at Galactic latitude 43\fdg70 and longitude 228\fdg91. 

In this paper, we present a study of the ionospheric phase delays for pulses emitted by pulsar B0950+08 over 1\,h obtained via observations using the {\it RadioAstron} ground-space radio interferometer. We analyse  the phase difference  between the antennas  based on the behaviour of cross-spectra responses of  large ground-space baselines. We describe a new technique for measuring the ionospheric TEC  and carry out an analysis of the derived ionospheric plasma based on the intercontinental distance between the Arecibo and WSRT stations. We show, that use of ground-space interferometer greatly simplifies the procedure of determining the ionospheric TEC  towards to the pulsar at the latitude and longitude of the ground station. 

In Section~\ref{sec:observations} we describe the observations and data processing. A detailed description of  phase incursion in a model simulating the development of the phase difference between the antennas is presented in Section~\ref{sec:measuring}. In Section~\ref{sec:ionospheric}, we consider the TEC measurements from the International Reference Ionosphere (IRI) project at the latitude and longitude of the Arecibo (AR) and Westerborg Synthesis Radio Telescope (WSRT) stations over the period of observations. In Section~\ref{sec:vtec}, we present a technique for measuring the ionospheric TEC at the ground station and analyse the preliminary results. Our conclusions are summarised in Section~\ref{sec:conclusions}.
\section{Observations and data processing}
\label{sec:observations}
\subsection{Space VLBI observations}
Pulsar observations were carried out on 25$^{th}$ January, 2012 between 05 and 06 UT, using a Space Radio Telescope (SRT) mounted on a Spectr-R satellite in conjunction with the Arecibo (latitude=18\fdg3442\,N and longitude=66\fdg7528\,W) and WSRT (latitude=52\fdg9147\,N and longitude=6\fdg6033\,E) stations. Two polarisation channels (Right Circular Polarisation, RCP, and Left Circular Polarisation, LPC) for the 16\,MHz upper sideband (USB) were  involved in this session. The central frequency of the receiver was $\nu_{\circ}=324$\,MHz. Data were sampled at the Nyquist frequency (31.25\,ns sampling time), and were stored on computer disks as separate scans over a continuous period of 570\,s with a gap of 30\,s in the two-bit digitising mode for the ground telescopes and over 270\,s with gap of 30\,s in one-bit digitising mode for the SRT. The {\it RadioAstron} project code was RAFS12. The duration of a continuous scientific session with SRT was limited by the thermal conditions of the on-board data transmission system and depended on the orientation of the satellite relative to the Sun. The observations were conducted at geocentric distance of between 291,000 and 292,000\,km near the apogee (317,000\,km). Scientific data from the SRT were transmitted in real time to the Pushchino tracking station (TS), and were recorded on a disk in RadioAstron Data Recorded (RDR) format \citep{2015CosRe..55..182A}. At the ground stations, the data were recorded in Mark5B and MKIV1\_4  VLBI data formats. To decode the data, we used the {\it mark5access} routine developed by Walter Brisken  from the DiFX software package \citep{2007PASP..119..318D}. The pulsar observations were planned using  the {\it FakeRat} software package \citep{2015CosRe..53..216Z}, which was developed for scheduling space VLBI observations using baselines exceeding the Earth's diameter. {\it FakeRat} software was designed to  model the position of a spacecraft (SC) with respect to an object, with regard to the design constraints on the orientation of the SC, the  angles of visibility of the SC by the TS (the Pushchino RT-22 and Green Bank 140 Foot Telescope), and the angles of visibility of a source tracked by the ground-bases  radio telescope. 
\subsection{SRT single-dish data processing}
The highly elliptical orbit of a SC evolves with time under the influence of the Moon and other external factors, which may be small but act over a long period. In the {\it FakeRat} model, the motion of the SC is defined by the predicted values of coordinates and velocities, which are calculated until the middle of 2019 \citep{2015CosRe..53..216Z}. However, the maximum possible accuracy of the ephemerides is required in order to obtain the responses of the interferometer. The values for each session were calculated at the Keldysh Institute of Applied Mathematics (IAM) of the Russian Academy of Sciences \citep{2014CosRe..52..342Z}. The accuracy of the reconstructed orbit is achieved by measurements of range and range rate in the C-band  at two Russian Command and Telemetry Stations at Bear Lakes (near Moscow) and at Ussurijsk, and Doppler measurements in the  X- and Ku-bands and optical measurements of the position of the SC in its orbit \citep{2013ARep...57..153K, 2014CosRe..52..326K}. The theoretical model provides an accuracy of 200\,m in the position, corresponding to delay errors of 0.7\,$\mu$s. Real measurements of the position of the SC can exceed the theoretical accuracy, and the clock delay for the SC can reach values of up to 10\,$\mu$s.  The delays for certain ground stations are not known with an adequate accuracy. An  initial fringe search should therefore be performed using a wider correlation window, with at least 2048 sampling periods of ($\pm 64 \mu$s) \citep{2017JAI.....650004L}.

The synchronisation of clocks between the SC and TS is done at the beginning of each scan by solving the light cone equation:
\begin{eqnarray}
\delta\tau_{srt}=\frac{1}{c}\{\vec{\boldsymbol r}_{srt}(\tau_{ts}+\delta\tau_{srt})-\vec{\boldsymbol r}_{ts}(\tau_{ts})\}
\label{eq:1}
\end{eqnarray}
where $\vec{\boldsymbol r}_{srt}$ is the position of the SC in the geocentric coordinate system according to its ephemerides, $\vec{\boldsymbol r}_{ts}$ is the position of the TS in the geocentric coordinate system, $\tau_{ts}$ is the time coordinate of the TS, $\delta\tau_{srt}$ is the time coordinate correction, and $c$ is the speed of light. 

During the observations, data are synchronised by the on-board hydrogen maser (H-maser) and transmitted to the TS. The start of a new observing session is controlled by the local H-maser. The time offset between the ground station and the on-board clock can be established by restoring the time delay between the SC and TS from a ballistic prediction of the  position of the SC \citep{2017JAI.....650004L}.
\subsection{VLBI data processing}
Data correlation is performed  using the ASC software FX correlator developed in the  Astro Space Center of the P.N.\,Lebedev Physical Institute \citep{2017JAI.....650004L} based on a pulse window, with compensation for the dispersion of the signal over the receiver band \citep{1975MComP..14...55H}. The ASC correlator uses the  {\it ORBITA2012} delay model, which is  based on {\it ARIADNA} algorithms. {\it ORBITA2012} calculates the delays between individual telescopes, including the SRT relative to the centre of the Earth. {\it ARIADNA} calculates the delays between the telescopes  using a "consensus model" \, (see Chap.\,11 in \citet{IERS}). This "consensus model" \, defines the standard reference system implemented by the International Earth Rotation and Reference System Service (IERS) and the models and procedures used for this purpose.

Complex cross-polarisation spectra, including auto-correlation spectra for each antenna, are formed between pairs of telescopes for all scans. Data were correlated using on- and off-pulse windows with 512 frequency channels. The on-pulse window was centred at the maximum of the average profile, while the off-pulse window used to record the noise was shifted by a quarter of the pulsar period later than the pulse profile, in order to avoid low-level precursor signals. The width of each window was $T_\circ=20.6$\,ms (about 10\% of the maximum of the average profile \citealt{2005AJ....129.1993M}). At the correlator output, we obtained spectra in standard Flexible Image Transport System (FITS) data format. The analysis of the correlated data was carried out with the help of the CFITSIO package \citep{1999ASPC...172..487P}.

The on- and off-pulse windows were moved synchronically with the pulsar period, and the longitudes of the on- and off-pulse windows with respect to the centre of Earth were determined using a new pulsar-timing package called TEMPO-2 \citep{2006MNRAS.369..655H} in the topocentric coordinate system. The timing ephemerides were taken from the ATNF pulsar catalogue. The longitude of the pulse maximum was preliminarily determined by us based on an analysis of the auto-correlation for Arecibo, the largest ground radio telescope. 

Post-correlation data reduction was performed using the Matplotlib plotting library and software specially developed by the authors.
\begin{figure}
    \includegraphics[width=\columnwidth]{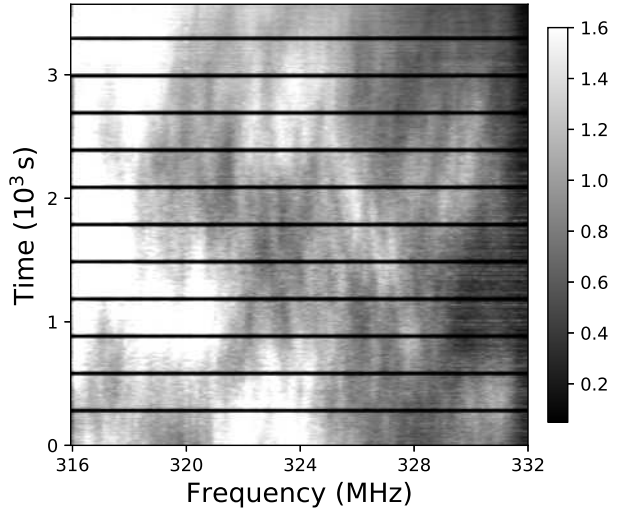}
    \caption{Dynamic cross-spectrum of PSR B0950+08 at an observing frequency of 324\,MHz for the Arecibo-WSRT ground baseline}.
    \label{fig:1}
\end{figure}
\begin{figure}
    \includegraphics[width=\columnwidth]{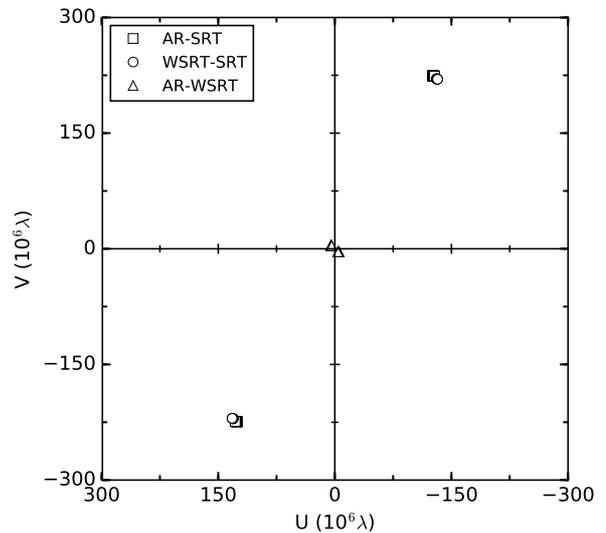}
    \caption{UV-coverage (visibility coverage of the Fourier domain) obtained for {\it RadioAstron} observations of PSR B0950+08 in units of spatial frequency, defined as the projected baseline in wavelengths $10^6\lambda$.}
    \label{fig:2}
\end{figure}
\section{Phase incursion}
\label{sec:measuring} 
\subsection{Phase coherence}
We obtained  dynamic cross-spectra for on-pulse scans between all possible pairs of telescopes showing diffraction scintillations in the time-frequency plane. Off-pulse spectra were used to correct for the receiver bandpass \citep{2013MNRAS.430.2815Z}.  Figure~\ref{fig:1} shows the dynamic cross-spectrum of PSR B0950+08 for the Arecibo-WSRT ground baseline. This spectrum was formed by summing cross-spectra with two opposite polarisations. The dark 30-second horizontal bands along the frequency axis show interruptions where the signal was not recorded. We can see a large-scale drift of diffraction patterns at a speed of $df/dt=4$\,MHz$/1000$\,s. This drift shows sloping scintillation structures defined by angular refraction from the "cosmic prism" . Following \citep{2017MNRAS.468.3709S}, we consider a phase difference between antennas $\Delta\Psi(\vec{\boldsymbol{r}}_1,\vec{\boldsymbol{r}}_2;\Delta\nu,t)$ as a combination of interstellar and ionospheric refractions at an intermediate frequency $\Delta\nu=\nu-\nu_{\circ}$:
\begin{eqnarray}
\Delta\Psi(\vec{\boldsymbol{r}}_1,\vec{\boldsymbol{r}}_2;\Delta\nu,t)=
\Delta \Psi_{int}(\vec{\boldsymbol{r}}_1,\vec{\boldsymbol{r}}_2;\Delta\nu)+
\Delta \Psi_{ion}(\vec{\boldsymbol{r}}_1,\vec{\boldsymbol{r}}_2;\Delta\nu,t)
\label{eq:22}
\end{eqnarray}

We parameterise the interstellar component using the interstellar refraction angle $\vec{\boldsymbol{\theta}}_{\circ}$ at the frequency $\nu_{\circ}$:
\begin{eqnarray}
\Delta \Psi_{int}(\vec{\boldsymbol{r}}_1,\vec{\boldsymbol{r}}_2;\nu_{\circ}) =
k(\vec{\boldsymbol{r}}_1-\vec{\boldsymbol{r}}_2)\cdot\vec{\boldsymbol{\theta}}_{\circ}
\label{eq:2}
\end{eqnarray}
where $k=2\pi/\lambda$ is the wavenumber, and $\lambda=c/\nu_{\circ}$ is the wavelength. Figure~\ref{fig:2} shows the resulting UV-coverage of the PSR B0959+08 observations. We see that the projections of the Arecibo-SRT and WSRT-SRT ground-space baselines do not change significantly, and we consider them fixed. For a fixed baseline $|\vec{\boldsymbol{r}}_1-\vec{\boldsymbol{r}}_2|$ the interstellar refraction depends only on $\Delta\nu$:
\begin{eqnarray}
\Delta \Psi_{int}(\vec{\boldsymbol{r}}_1,\vec{\boldsymbol{r}}_2;\Delta \nu) = 
\frac{\Delta \nu}{\nu_{\circ}} \Delta \Psi_{int}(\vec{\boldsymbol{r}}_1,\vec{\boldsymbol{r}}_2;\nu_{\circ}) 
\label{eq:3}
\end{eqnarray}

We examine the ionospheric component at separate small time intervals $n$ from $(t_{n,\circ}-T/2)$ to $(t_{n,\circ}+T/2)$, where $t_{n,\circ}$ is the central moment of the $n$-th time interval. The duration of time interval  is determined by a number of accumulated spectra as:
\begin{eqnarray}
  T=NT_\circ
\end{eqnarray}
where $N=50,\,200$ and 330 are the numbers of accumulated spectra for the Arecibo-WSRT, Arecibo-SRT and WSRT-SRT baselines, respectively. We then obtain an expression for the ionospheric component at $\Delta\nu$:
\begin{eqnarray}
\Delta \Psi_{ion}(\vec{\boldsymbol{r}}_1,\vec{\boldsymbol{r}}_2;\Delta\nu,t_{n,m}) = \frac{\Delta\nu}{\nu_{\circ}}
\Delta\Psi_{ion}(\vec{\boldsymbol{r}}_1,\vec{\boldsymbol{r}}_2;\nu_{\circ},t_{n,m})
\label{eq:4}
\end{eqnarray}
where $\Delta\Psi_{ion}(\vec{\boldsymbol{r}}_1,\vec{\boldsymbol{r}}_2;\nu_{\circ},t_{n,m})$ is the ionospheric component at $\nu_{\circ}$, $t_{n,m}$ is the particular pulse time $m$ within the time interval $n$. 

Furthermore, $\Delta\Psi_{ion}(\vec{\boldsymbol{r}}_1,\vec{\boldsymbol{r}}_2;\nu_{\circ},t_{n,m})$ 
as a time function can be expanded around the central moment $t_{n,\circ}$ by a 
Taylor series up to the linear term:
\begin{eqnarray}
\Delta \Psi_{ion}(\vec{\boldsymbol{r}}_1,\vec{\boldsymbol{r}}_2;\nu_{\circ},t_{n,m}) = \nonumber \\
\Delta \Psi_{ion}(\vec{\boldsymbol{r}}_1,\vec{\boldsymbol{r}}_2;\nu_{\circ},t_{n,\circ})+
\frac{(t_{n,m}-t_{n,\circ})}{T}\Phi_{ion}(\vec{\boldsymbol{r}}_1,\vec{\boldsymbol{r}}_2;\nu_{\circ},t_{n,\circ})
\label{eq:5}
\end{eqnarray}
where $\Phi_{ion}(\vec{\boldsymbol{r}}_1,\vec{\boldsymbol{r}}_2;\nu_{\circ},t_{n,\circ})=T\frac{d}{dt}[\Delta\Psi_{ion}(\vec{\boldsymbol{r}}_1,\vec{\boldsymbol{r}}_2;\nu_{\circ},t)]_{|t=t_{n,\circ}}$ is the phase incursion.

The phase difference between antennas can be represented by the second moment $SM_n(\nu)$, averaging over time and frequency as:
\begin{eqnarray}
  SM_n(\nu)=\langle J(\Delta\nu,t_{n,m})J^*(\Delta\nu+\nu,t_{n,m})\rangle_{t_{n,m}}= \nonumber\\  \langle j(\Delta\nu,t_{n,m})j^*(\Delta\nu+\nu,t_{n,m})\rangle_{t_{n,m}}\varphi_n(\nu)
\end{eqnarray}
where $j(\Delta\nu,t_{n,m})$ are coefficients of the interferometric response, $\nu$ is the frequency shift, and $\varphi_n(\nu)$ is defined by $\Delta\Psi(\vec{\boldsymbol{r}}_1,\vec{\boldsymbol{r}}_2;\Delta \nu,t_{n,\circ})$ from Equation~\ref{eq:22} taking into account Equations~\ref{eq:3}--\ref{eq:5}:
\begin{eqnarray}
&&  \varphi_n(\nu)= 
  \bigl\langle \exp\{-i[\Delta\Psi(\vec{\boldsymbol{r}}_1,\vec{\boldsymbol{r}}_2;\Delta\nu,t_{n,m}) \nonumber \\
&&\qquad\qquad\qquad  -\Delta\Psi(\vec{\boldsymbol{r}}_1,\vec{\boldsymbol{r}}_2;\Delta\nu+\nu,t_{n,m})]\}\bigr\rangle_{t_{n,m}} = \nonumber\\
&& \exp\{-i(\nu/\nu_{\circ})[\Delta\Psi_{int}(\vec{\boldsymbol{r}}_1,\vec{\boldsymbol{r}}_2;\nu_{\circ})+
\Delta\Psi_{ion}(\vec{\boldsymbol{r}}_1,\vec{\boldsymbol{r}}_2;\nu_{\circ},t_{n,\circ})]\} \nonumber \\
&&\qquad\qquad \times sinc[(\nu/2\nu_{\circ})\Phi_{ion}(\vec{\boldsymbol{r}}_1,\vec{\boldsymbol{r}}_2;\nu_{\circ},t_{n,\circ})]
\label{eq:6}
\end{eqnarray}
\begin{figure*}
    \begin{center}
	\includegraphics[width=\textwidth]{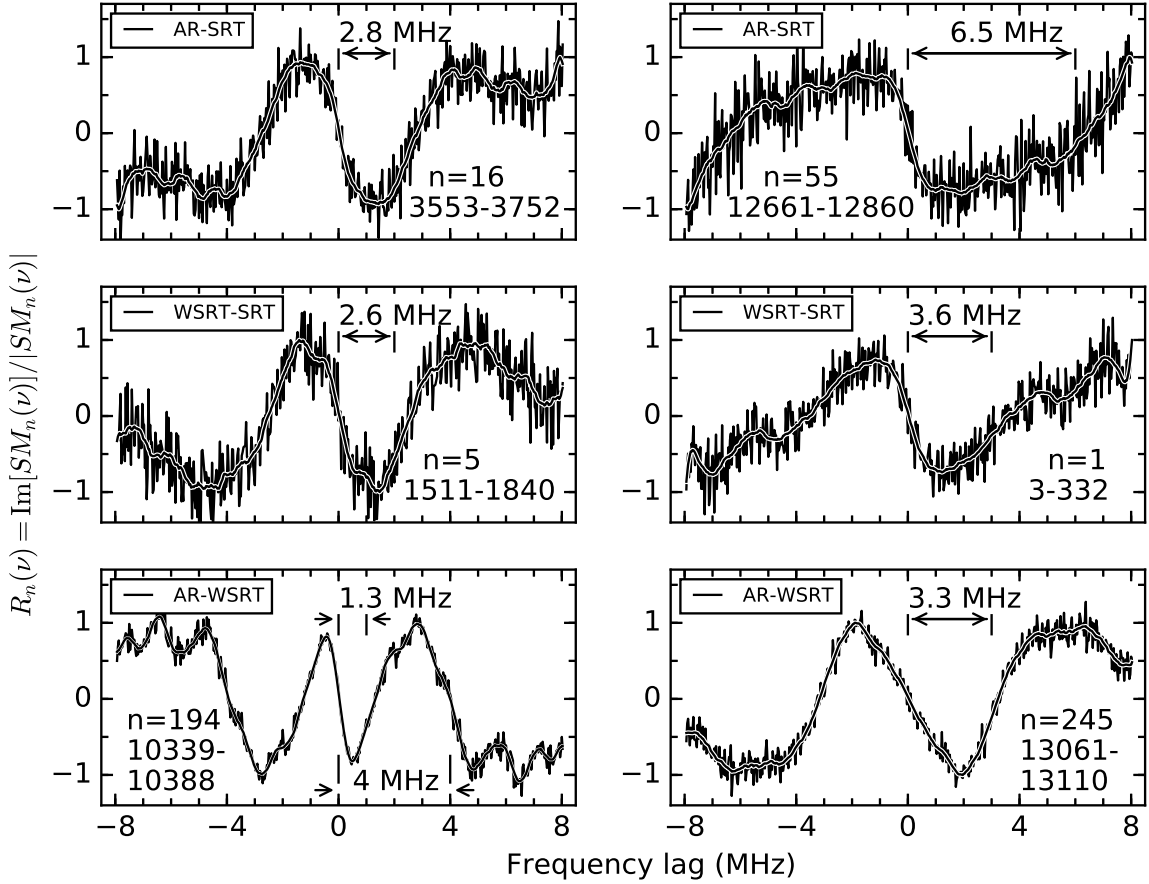}
	\caption{$R_n(\nu)$ is the ratio of the imaginary part $SM_n(\nu)$ to the modulus $|SM_n(\nu)|$  for the Arecibo-SRT (upper panel), WSRT-SRT (middle-panel) and Arecibo-WSRT baselines (lower panel) at varying time. In each panel, the serial number of the averaging interval  $n$ and the numbers of averaged pulses are shown. The pulse numbers are defined as $n^{\prime}+(n-1)N\div n^{\prime}+iN-1$, where $n^{\prime}$ is a pulse offset. The dependencies are smoothed using a fourth-order Savitzky-Golay filter. The numerical value of the "zero factor"\ $\nu^*$ is presented in MHz. Examples with small values of $\nu^*$ are shown in the left-hand panels, while those with large values are shown in the right-hand panels. Example with two "zero factors"\  is shown in the lower left-hand panel (for more details, see the discussion in Section~\ref{sec:zero}).}
	\label{fig:3}
    \end{center}
\end{figure*}

\begin{figure}
    \includegraphics[width=\columnwidth]{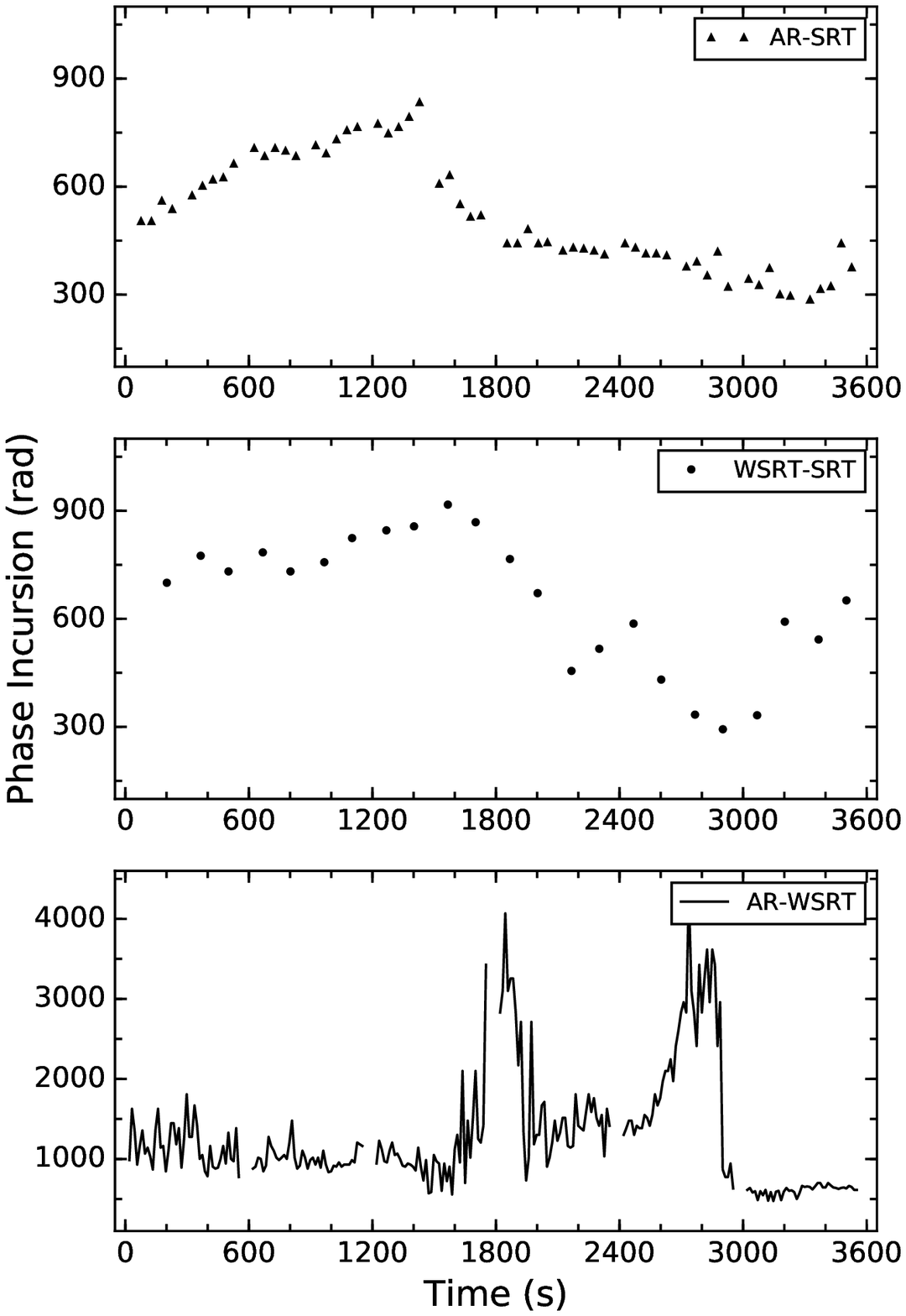}
    \caption{Phase incursion as a function of time for the Arecibo-SRT (upper panel), WSRT-SRT (middle panel) and Arecibo-WSRT (lower panel) baselines.}
    \label{fig:4}
\end{figure}
Let us define the ratio of the imaginary part of the second moment to its modulus as $R_n(\nu)$. Figure~\ref{fig:3}  shows examples of $R_n(\nu)$  for the ground-space and ground baselines, averaged over different time intervals.
\begin{itemize}
\item The value of $R_n(\nu)$ is zero at $\nu=0$, as determined by the $\sin\{-(\nu/\nu_{\circ})[\Delta\Psi_{int}(\vec{\boldsymbol{r}}_1,\vec{\boldsymbol{r}}_2;\nu_{\circ})+\Delta\Psi_{ion}(\vec{\boldsymbol{r}}_1,\vec{\boldsymbol{r}}_2;\nu_{\circ},t_{n,\circ})]\}$ factor in Equation~\ref{eq:6}.
\item The other zero value of $R_n(\nu)$  is determined at $\nu=\pm\nu^*(t_{n,\circ})$ by the $sinc[(\nu/2\nu_{\circ})\Phi_{ion}(\vec{\boldsymbol{r}}_1,\vec{\boldsymbol{r}}_2;\nu_{\circ},t_{n,\circ})]$ factor. We designate this frequency as the "zero factor"\ . It is defined by the argument of $sinc$ equal to $\pm\pi$.
\end{itemize}
\subsection{The "zero factor"}
\label{sec:zero}
In addition, to increase the signal-to-noise ratio without destroying the signal itself, we apply a fourth-order Savitzky-Golay filter \citep{1964AnaCh..36.1627S} along the frequency axis. As can be seen from Figure~\ref{fig:3}, the filtering completely smooths out the small-scale noise fluctuations that remain after averaging.

Figure~\ref{fig:3} shows $R_n(\nu)$ with typical small values of $\nu^*$ in the left-hand panels and with typical large values in the right-hand panels. The values of $\nu^*$ for short ground baselines occupy a frequency domain from 0 to 4\,MHz, while $\nu^*$ for the large ground-space baselines shifts the frequency domain  to large values of between  2.5 and 8\,MHz. 

We note that for both ground-space baselines  and in most cases for the ground baseline, we have $R_n(\nu)$ with only one \mbox{"zero factor"\ }, while for the ground baseline only, we detect $R_n(\nu)$ with two \mbox{"zero factors"\ } over a few averaging intervals. An example with two  \mbox{"zero factors"\ } is shown in Figure~\ref{fig:3}, lower left-hand image. The first \mbox{"zero factor"\ } is found at  1.3\,MHz, and the second is found at 4.0\,MHz. This  behaviour of $R_n(\nu)$ is easy to explain using our phase model (see Equation~\ref{eq:6}). When the value of the first "zero factor"\ is sufficiently small (in our example, $\nu^*=1.3$\,MHz), an additional "zero factor"\ (in our example, $\nu^{**}=4.0$\,MHz) may appear within the band of the frequency lag, ($0-8$)\,MHz. The second \mbox{"zero factor"\ }is defined by an argument of $sinc$ equal to $2\pi$. For clarity, we used only the first value for the "zero factor"\, regardless of whether or not a second value was detected.

We can see that the number of detected "zero factors"\,  depends on the state of the ionospheric plasma, the baseline of the interferometer and the band pass receiver. 
\subsection{Phase incursion}
Finally, for our values of $\nu^*(t_{n,\circ})$ we find the phase incursion as function of time:
\begin{eqnarray}
\Phi_{ion}(\vec{\boldsymbol{r}}_1,\vec{\boldsymbol{r}}_2;\nu_{\circ},t_{n,\circ})=2\pi\nu_{\circ}/\nu^*(t_{n,\circ})
\label{eq:7}
\end{eqnarray}

Figure~\ref{fig:4} shows values of $\Phi_{ion}(\vec{\boldsymbol{r}}_1,\vec{\boldsymbol{r}}_2;\nu_{\circ},t_{n,\circ})$ for the ground-space and ground baselines. We see that for the ground baseline, the ionospheric phase is larger and varies more rapidly with frequency than for the ground-space baselines, and we  draw attention to the significant differences between the time dependencies of the phase incursions for the ground-space and ground baselines, as follows:
\begin{enumerate}
\item This is due to a significant increase in $\Phi_{ion}(\vec{\boldsymbol{r}}_1,\vec{\boldsymbol{r}}_2;\nu_{\circ},t_{n,\circ})$ during the first 30 min of the observations for both ground-space baselines, which is absent from the ground baseline. 
\item Sudden jumps in  $\Phi_{ion}(\vec{\boldsymbol{r}}_1,\vec{\boldsymbol{r}}_2;\nu_{\circ},t_{n,\circ})$ to long lag times are then detected at the ground baseline, which are absent from  the ground-space baselines.
\item Despite certain similarities in the time dependencies of $\Phi_{ion}(\vec{\boldsymbol{r}}_1,\vec{\boldsymbol{r}}_2;\nu_{\circ},t_{n,\circ})$ on the ground-space baselines, differences definitely exist.
\end{enumerate}
These issues are discussed in more detail in Section~\ref{sec:vtec}.
\section{Ionospheric information derived from IRI measurements}
\label{sec:ionospheric}
\begin{figure}
    \includegraphics[width=\columnwidth]{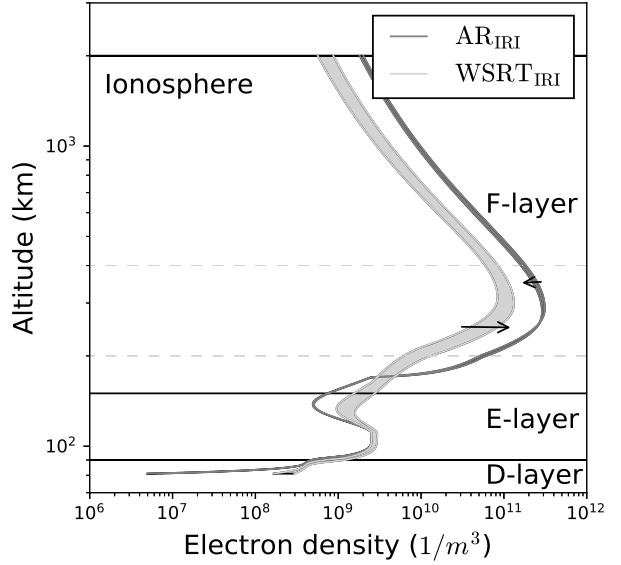}
    \caption{Profiles of ionospheric electron density from IRI at 05-06 UT on 25$^{th}$ January, 2012 at the positions of the Arecibo and WSRT stations. The arrows indicate the directions along which the profile changes. The ionospheric layers are shown. The ionospheric profile extends up to a height of 2,000\,km.}
    \label{fig:5}
\end{figure}

Here we present an estimate of the TEC obtained from other independent measurements. The profile of the ionospheric electron density over height can be taken from the IRI-2018 model  \citep{2018AdRS...16....1B}. Figure~\ref{fig:5} shows two profiles: one for Arecibo, and the second for WSRT. The arrows indicate the directions along which the profiles change during the session. Using the available data, we estimate the vertical TEC (VTEC) from IRI as follows:
\begin{eqnarray}
\mbox{VTEC}=\sum_{h=60}^{2,000}\mbox{VTEC}_h 
\end{eqnarray}
where $\mbox{VTEC}_h$ is the value of the VTEC for an individual layer at height $h$ with a thickness of 1\,km. 

Excessive values of the slant TEC (STEC) along the line of sight can be obtained from the thin-layer model, which approximates the ionosphere as a thin layer of a spherical shell at a fixed height $H$ from the Earth's surface. Then, based on the VTEC and the zenith angle of the ray path passing through this thin layer, the STEC can be represented as:
\begin{eqnarray}
\mbox{STEC}=\mbox{VTEC}\sec{\theta}
\label{eq:8}
\end{eqnarray}
where $\theta$ is the zenith angle at the Ionospheric Pierce Points (IPP). The zenith angle of the ray at the bottom of the ionosphere  is connected with the zenith angle at the observer by the sine law \citep[e.g.,][]{2017isra.book.....T}:
\begin{eqnarray}
\theta=\arcsin^{-1}\Bigl(\frac{R_{\earth}}{R_{\earth}+H}\sin\theta^{\prime}\Bigl)
\label{eq:80}
\end{eqnarray}
where $R_{\earth}=6,371$ km is the mean radius of the Earth. Therefore, from Equation~\ref{eq:8}:
\begin{eqnarray}
\mbox{STEC}= \mbox{VTEC}\times \Bigl[1-\Bigl(\frac{R_{\earth}}{R_{\earth}+H}\sin\theta'\Bigr)^2\Bigr]^{-1/2}
\label{eq:70}
\end{eqnarray}

\begin{figure}
    \includegraphics[width=\columnwidth]{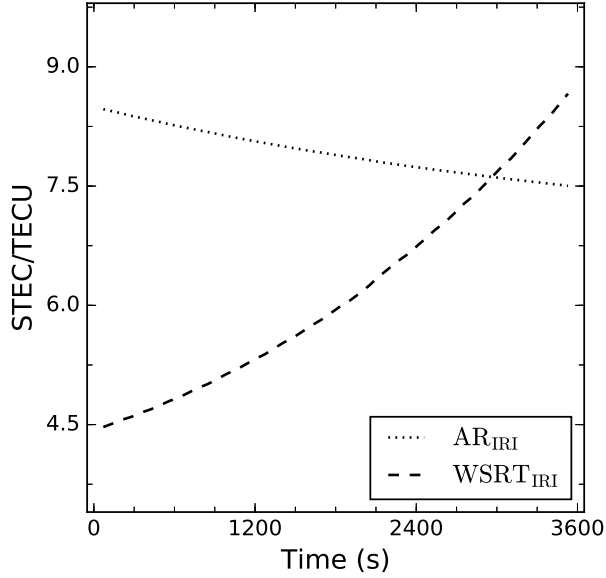}
    \caption{Variation in the ionospheric IRI STEC for the Arecibo and WSRT stations during the session, along the line of sight to PSR B0950+08.}
    \label{fig:6}
\end{figure}

We present the STEC values approximated by the thin layer model at $H=450$\,km. Figure~\ref{fig:6} shows the IRI STEC for the Arecibo and WSRT stations in the direction of PSR B0950+08 as function of time. We can see that approximately 50 minutes after the beginning of the session, the STEC for  Arecibo and WSRT became equal.

Note that equality of the STEC at the ends of the interferometric baseline for the ground stations follows indirectly from our analysis. In the second half of the observations for the ground interferometer (see Figure~\ref{fig:4}, lower panel) sudden jumps appear in the phase  incursion $\Phi_{ion}(\vec{\boldsymbol{r}}_1,\vec{\boldsymbol{r}}_2;\nu_{\circ},t_{n,\circ})$, while there are no such jumps for the ground-space interferometers. In our opinion, these jumps occur every time the constant term $\Delta \Psi_{ion}(\vec{\boldsymbol{r}}_1,\vec{\boldsymbol{r}}_2;\nu_{\circ},t_{n,\circ})$ is zeroed out in the Taylor expansion (see Equation~\ref{eq:5}). These events happen each time the STEC is aligned at the ends of the interferometric baseline. In this case, only one term (the first derivative) remains in the Taylor expansion, and the phase incursion model no longer works correctly. According to our analysis, zeroing of the constant term in the Taylor expansion occurs twice during the session: the first time, sudden jump is detected about 30 min after the beginning of the observations, while the second time, a similar event occurs about 15 min later. Thus, in order to implement correct data processing, it is necessary to introduce at least the second derivative into Equation~\ref{eq:5}, and we intend to address this in future work. In Section~\ref{sec:vtec} below, our consideration of the ionosphere will be limited by the Arecibo-SRT and WSRT-SRT ground-space baselines, as discussed in Section~\ref{sec:introduction}. Along the large ground-space baseline, the ionospheric effects are detected only for the ground stations, so the implementation of the case considered above (equality of the STEC at the ends of the interferometric ground-space baseline) is physically impossible.
\section{Results for space VLBI and comparison with IRI data}
\label{sec:vtec}
We represent the phase difference $\Phi_{ion}(\vec{\boldsymbol{r}}_1,\vec{\boldsymbol{r}}_2;\Delta\nu,t_{n,\circ})$ between two phase incursions $\Phi_{ion}(\vec{\boldsymbol{r}}_1,\vec{\boldsymbol{r}}_2;\nu,t_{n,\circ})$ measured at frequencies $\nu_k$ and $\nu_{\circ}$ ($\delta\nu=|\nu_k-\nu_{\circ}|\leqslant 8$\,MHz) within the time interval $[t_{n,\circ}-T/2, t_{n,\circ}+T/2]$ as:
\begin{eqnarray}
\Phi_{ion}(\vec{\boldsymbol{r}}_1,\vec{\boldsymbol{r}}_2;\delta\nu,t_{n,\circ})=
\Phi_{ion}(\vec{\boldsymbol{r}}_1,\vec{\boldsymbol{r}}_2;\nu_{\circ},t_{n,\circ})
\frac{\delta\nu}{\nu_{\circ}}
\label{eq:11}
\end{eqnarray}
The phase delay for a cold, collisionless plasma with no magnetic field for a frequency exceeding the plasma frequency $\nu_p$ can be approximated by the following expression \citep[e.g.,][]{2009A&A...501.1185I}:
\begin{eqnarray}
\varphi \simeq \frac{e^2}{4\pi\varepsilon_{\circ}m_ec\nu}\int n_e dl=8.447\times 10^3\frac{1}{\nu}\Bigl(\frac{\mathrm{STEC}}{\mathrm{TECU}}\Bigl)
\label{eq:12}
\end{eqnarray}
where $e$ is the electron charge, $\varepsilon_{\circ}$ is the vacuum permittivity, and $m_e$ is the electron mass. The integral over $n_e$ is the STEC along the line of sight. The plasma frequency is 
\begin{eqnarray}
\nu_p=\frac{e}{2\pi}\sqrt{\frac{n_e}{\varepsilon_{\circ}m_e}}
\label{eq:20}
\end{eqnarray}
Typical values for the plasma frequency for the ionosphere are within the range  1 to 10\,MHz. Cosmic radiation at lower than the plasma frequency does not reach the Earth's surface.

The same phase difference $\Phi_{ion}(\vec{\boldsymbol{r}}_1,\vec{\boldsymbol{r}}_2;\delta\nu,t_{n,\circ})$ can be obtained based on the STEC  at the same $\delta\nu$:
\begin{eqnarray}
\Phi_{ion}(\vec{\boldsymbol{r}}_1,\vec{\boldsymbol{r}}_2;\delta\nu,t_{n,\circ})=8.447\times 10^3\Bigl(\frac{\mathrm{STEC}}{\mathrm{TECU}}\Bigl)\frac{\delta\nu}{\nu^2_{\circ}}
\label{eq:13}
\end{eqnarray}
Consequently, taking into account the thin layer model of the ionosphere in Equation~\ref{eq:70}, the value of the VTEC for the ground arm of the ground-space interferometer is given by the relation:
\begin{eqnarray}
\frac{\mathrm{VTEC}}{\mathrm{TECU}}= 1.18 \cdotp 10^{-4}\nu_{\circ}
\Phi_{ion}(\vec{\boldsymbol{r}}_1,\vec{\boldsymbol{r}}_2;\nu_{\circ},t_{n,\circ}) \nonumber \\
\times\sqrt{1-\Bigl(\frac{R_{\earth}}{R_{\earth}+H}\sin\theta'(t)\Bigr)^2}
\label{eq:14}
\end{eqnarray}
\begin{figure}
    \includegraphics[width=\columnwidth]{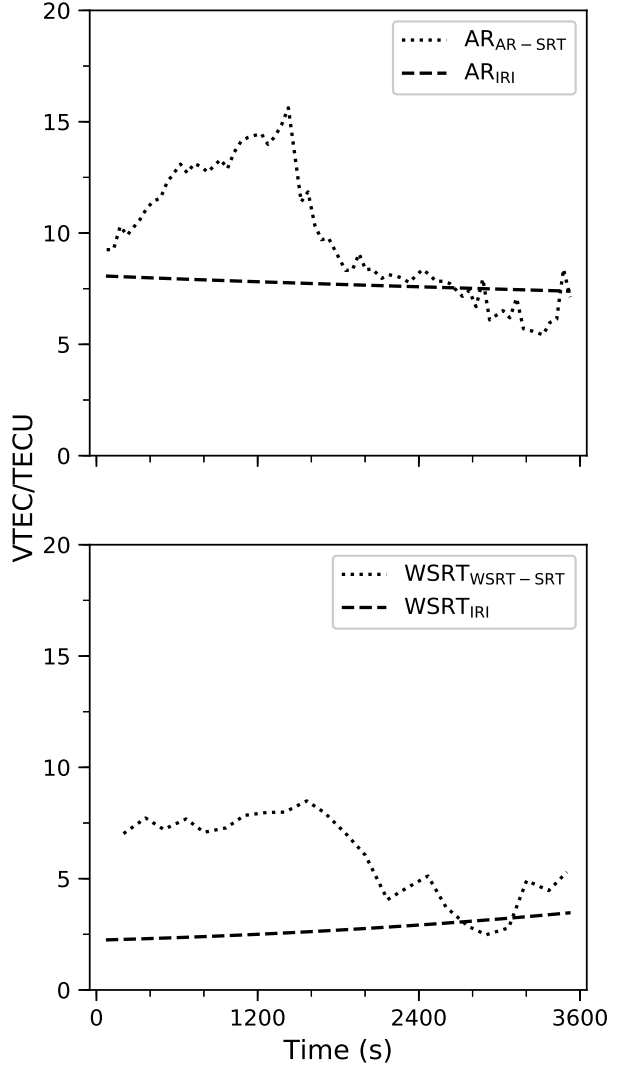}
    \caption{VTEC for Arecibo (upper panel) and WSRT (lower panel) as a function of time. The corresponding values of the IRI VTEC are also shown.}
    \label{fig:7}
\end{figure}
Figure~\ref{fig:7} represents the VTEC for the Arecibo (upper panel) and WSRT (lower panel) stations as function of time. The time resolution for the VTEC is 0.8 and 1.4\,min for the Arecibo and WSRT stations, respectively. For comparison, we show the corresponding IRI VTEC, as follows:
\begin{itemize}
\item  At both ground stations, during the first 30--40 min after the beginning of the session, large half-hour structures in VTEC are detected in comparison with the VTEC from IRI. Moreover, the VTEC for the Arecibo station shows a increase to the level of 15.5\,TECU after  $\sim25$\,min after the beginning of the observations, while the VTEC for the WSRT remains at a value of $7.37\pm0.71$\,TECU within the first 30\,min, without significant fluctuations.
\item In the second half of the session, VTEC decreases to a level corresponding to the VTEC from IRI. This decrease occurs along different paths for the Arecibo and WSRT stations. From Figure~\ref{fig:7}, we find that the VTEC for Arecibo decreases exponentially and lasts $\sim 10$\,min, while the VTEC for WSRT decreases almost linearly and lasts twice as long. 
\item At the low geomagnetic latitude of Arecibo (27\fdg 85\,N), we obtain the expected values for the  VTEC, which are approximately twice as large as the corresponding values at the mid-geomagnetic latitude of WSRT (53\fdg 48N).
\item The local time (LT) of the expected daily minimum VTEC depends on the diurnal and seasonal variation in VTEC and coordinates of of the observing station. According to published data \citep[see e.g.][]{1984RaSc...19..779S,2014AdSpR..54.1813M} the observed daily minimum VTEC occurs between approximately $3^h$ and $5^h$\,LT. The observations for the Arecibo station were conducted after sunset ($18^h16^m$ LT) between $1^h$ and $2^h$ LT and those for the  WSRT before sunrise ($8^h27^m$ LT) between $6^h$ and $7^h$ LT. The expected daily variations are clearly shown  in the VTEC derived from IRI. Figure~\ref{fig:7} shows the daily decrease in IRI VTEC for the Arecibo station and the daily increase in the IRI VTEC for  WSRT. Unfortunately, the daily variations in the ionospheric VTEC derived from the ground-space interferometer are greatly obscured by the structures detected in this experiment.
\end{itemize}
According to a preliminary analysis, the detected structures may be due to the influence of interplanetary and magnetospheric events during the observations \citep{1997GMS....98...77T,2005P&SS...53..189Y}. The event (05:00 -- 05:30\,UT 25$^{th}$ January, 2012) was observed during disturbed interplanetary conditions, and fell into the compression region between the slow and fast solar wind streams (the so-called corotating interaction region, CIR), which began with a forward interplanetary shock at $\sim 15$\,UT 24$^{th}$ January, 2012 and ended with a reverse interplanetary shock at $\sim 17$\,UT 25$^{th}$ January, 2012. The front edge of the CIR collided with the rear edge of the previous magnetic cloud and was therefore additionally compressed (see the catalogue of large-scale solar wind phenomena\footnote{ftp://ftp.iki.rssi.ru/pub/omni/} in \citealt{2009CosRe..47...81Y}). Ground magnetic stations showed that just after the shock at  $\sim 15$\,UT 24$^{th}$ January, 2012, Storm Sudden Commencement (SSC) was observed on the Earth with a moderate two-stage magnetic storm with an intermediate minimum of Dst index -44\,nT at $\sim 20$\,UT on 24$^{th}$ January, 2012 and the main minimum in the Dst index was  -75\,nT at $\sim 11$\,UT 25$^{th}$ January, 2012 \footnote{http://wdc.kugi.kyoto-u.ac.jp/dst\_final/201201/index.html}. These changes in the magnetic field occurred as  a consequence of the global restructuring of the magnetosphere-ionosphere current system, and were are accompanied by large perturbations of the ionosphere \citep{2008JGRA..113.0A02M,2008GMS...181....9P,2017SSRv..212.1271K, 2018JSWSC...8A..33T}.
\section{Conclusions}
\label{sec:conclusions}
We carried out a successful {\it {RadioAstron}} VLBI measurements of the ionospheric electron concentration, based on ground-space interferometric responses of cross-spectra. This work became possible after launching the SRT and using it in conjunction with other ground antennas on large baselines of nearly 25 Earth diameters. Due to the fact that the orbital telescope was located at a large distance from the Earth's surface (above the ionosphere), only the Arecibo and WSRT ground stations could be used to record the ionospheric effects. Based on simulations of the interstellar and ionospheric refractions and the phase delay, we present a technique for processing and analysing the ionospheric TEC localised at the ground stations of the ground-space interferometer at 324\,MHz. Using this technique, we  obtained  the ionospheric TEC with a resolution of 0.8 and 1.4\,min for Arecibo and WSRT, respectively. An analysis of the detected TECs shows almost synchronous half-hour structures in the ionosphere at the intercontinental distance between the Arecibo and WSRT stations, which are approximately twice as large in amplitude as the corresponding values of the TEC derived from IRI. According to a preliminary analysis, these structures were observed during a geomagnetic storm with minimum Dst index $\sim 75$\,nT generated by interplanetary disturbances, and may be due to the influence of interplanetary and magnetospheric phenomena on ionospheric disturbances. The results presented in this paper may be useful in correcting or at least investigating the impact of the ionosphere on VLBI observations. 
\section{Future work}
From the middle of 2011 until the beginning of 2019 about 90 observations of pulsars were carried out  using {\it RadioAstron} in conjunction with different arrays of ground telescopes at two frequencies 324\,MHz (P band) and 1668\,MHz (L band). In this paper, we  have considered data only at 324\,MHz. We plan to test the robustness of our proposed technique by processing and analysing the TEC at a second standard {\it RadioAstron} frequency with the double side-band (DSB), each with two polarisations. The {\it RadioAstron} mission is also able to operate simultaneously on two frequency channels each receiving a radio signal with different  polarisations. These modes will allow us to expand the technique by including  two frequency channels at the same time under different ionospheric conditions and will facilitate obtaining more information regarding dynamic processes in the ionosphere.

\section*{Acknowledgements}
The RadioAstron project is led by the Astro Space Center of the Lebedev Physical Institute of the Russian Academy of Sciences and the Lavochkin Scientific and Production Association under a contract with the Russian Federal Space Agency, in collaboration with partner organizations in Russia and other countries. We are very grateful to the staff at the Arecibo Observatory, which is operated by SRI International under a cooperative agreement with the National Science Foundation (AST-1100968), and in alliance with Ana G. M\'{e}ndez-Universidad Metropolitana, and the Universities Space Research Association,  and the WSRT,  which is operated by ASTRON/NWO, for their support. The authors thank Leonid Petrov, NASA Goddard Space Flight Center for many interesting discussions and valuable suggestions. This research was supported by Russian Science Foundation, project 16-12-10062.
%%%%%%%%%%%%%%%%%%%%%%%%%%%%%%%%%%%%%%%%%%%%%%%%%%

%%%%%%%%%%%%%%%%%%%% REFERENCES %%%%%%%%%%%%%%%%%%

% The best way to enter references is to use BibTeX:

\bibliographystyle{mnras}
\bibliography{mybib} % if your bibtex file is called example.bib

% Don't change these lines
\bsp	% typesetting comment
\label{lastpage}
\end{document}